\documentstyle[preprint,aps]{revtex}
\begin{document} 
\tightenlines
\title{Stretching DNA: Role of electrostatic interactions} 
\author{Namkyung Lee and  D. Thirumalai}
\address{Institute  of Physical Science and Technology, 
University of Maryland,College Park, Maryland  20742} 
\date{\today} 
\maketitle 
 
\begin{abstract}
The effect of electrostatic interactions on the stretching of DNA is 
investigated using a simple worm like chain model. In the
limit of small force there are large conformational
fluctuations which are treated using a self-consistent variational approach.
For small values of the external force $f$, we find  theoreticlly 
and by a simple blob picture that the extension scales as $f r_D$ where 
$r_D$ is the Debye screening length.
In the limit of large force the electrostatic effects can be  accounted for within 
the semiflexible chain model of DNA by assuming that only small excursions from
rod-like conformations are possible. In this regime the extension 
approaches the contour length as $1/\sqrt{f}$ where $f$  is the magnitude of
the external force. The theory is used to analyze  
experiments that have measured the extension of double-stranded DNA subject  to
tension at various salt concentrations. The theory reproduces nearly 
quantitatively the elastic 
response of DNA at small and large values of $f$ and for all concentration of
the monovalent counterions. The limitations of the theory are also pointed out.

\end{abstract}

\section{INTRODUCTION}
The technical advances in the manipulation of single molecules has enabled  
the  probe of  mechanical and relaxational measurement in both equilibrium\cite{smith2,stick} 
and non-equilibrium conditions\cite{smith1}.   Among the first  of such experiments was  
the investigation  of the elastic response of $\lambda$ - bacteriophage DNA ($\lambda$ DNA)
 molecules subject to tension\cite{smith2}.   These experiments and others have offered 
a window into the behavior of biological  molecules on scales ranging from 
several nanometer to few microns.  They also provide an opportunity to 
understand the limits of validity of    theories based on generic polymer 
models such as Edwards model or simple model of semiflexible chain such as the worm like chain (WLC)  model\cite{doi&edwards}.

The earliest theories describing the elastic response of WLC  subject to tension, which appear to be 
relevant to the experiments of Smith et al.\cite{smith2}, were due to Fixman and Kovac \cite{fixman} and Crabb and Kovac\cite{crabb}. 
The precise  experiments made possible by nanomanipulation of single molecules
\cite{smith2,stick} 
has  demanded more accurate theories.  In these experiments the response to 
a constant force on a magnetic bead attached  to $\lambda$ DNA 
in a solution of varying salt concentration is used   to probe DNA elasticity.
After these pioneering experiments were reported  
several theoretical studies followed\cite{brochard,marko,ha}.  These papers showed   that, when the salt concentration is sufficiently large, then the simple WLC model 
subject to tension quantitatively reproduces the force-extension curves.
 In particular, the asymptotic approach to full extension  at large force,
was shown to follow the $f^{-1/2}$ law, where $f$ is the external force.

Since DNA is highly charged one expects ion effects to be extremely
important in determining the elasticity.  It has recently  been 
established that 
the elastic response and  persistence length  of DNA greatly depend 
not only on the salt concentration but also on the valence of counterions\cite{baumann}.
As  in the case of simple polyelectrolytes,    
systematic theories of DNA subject to tension is difficult
due to the presence of several competing length scales.  In particular, 
at intermediate values of the external force, the interplay of several forces,
namely intrinsic persistence length, electrostatic repulsion, counterion fluctuation, all conspire to determine the conformation of DNA.  

Marko and Siggia\cite{marko} included the effects of electrostatic interaction, within 
a WLC picture of DNA by replacing the intrinsic persistence length by 
a scale dependent effective persistence length\cite{barrat&joanny}.
 Such a description  
implicitly  assumes  that
 the intrinsic persistence length is large, and hence only 
small excursions  in DNA from rod-like conformations are  tolerated.  The interaction 
between the charges is  assumed to obey the Debye-H\"uckel potential, 
$V_{DH}(r) = l_B e^{-\kappa r}/r$.
The  ionic strength  of the solution  $I$ is related to the screening
length $r_D = \kappa^{-1}$ 
through  the relation $\kappa^2 = 4\pi l_B I $, and  
the Bjerrum length $l_B = e^2/(\epsilon k_B T)$ represents the  strength of interaction.
The scale dependent  effective persistence length $l_p^{eff}$  varies from 
 $l_o + l_{OSF}$ ( in length scale $r \gg r_D$ ) to $l_o$ ($r\ll r_D$) , where
 $l_{OSF} = l_B /4\kappa^2 A^2$ \cite{OSF1,OSF2} with  $A$ being the mean distance between charges.
Marko and Siggia  considered the limit when the applied force is so large
that  the angle fluctuation of tangent vector 
with respect to the direction of applied force is small which 
in consistent with   the assumption of the OSF theory\cite{OSF1,OSF2}.
The  chain extension $z$ in the presence of tension  approaches $L$  like  $f^{-1/2}$,  in good 
agreement with experiment at  large force.  
However, when the  conformation of DNA is coil-like ( at small force )  
the theory shows strong deviation from experimental results.

In order to probe the effects of electrostatic interactions at small 
values of $f$ and $I$  we suggest a mean field variational 
approach.   The mean field model  of semiflexible
chains \cite{ha&dave3,LNN}  successfully describes    
 the conformation of the  stretching DNA
 by a constant force at high salt concentration, where the electrostatic interactions are negligible\cite{ha}.
In this model, the hard constraint on  tangential vector $\bf u\rm^2(s) = 1$ is replaced by a global constraint $<\bf u\rm^2(s)> = 1$ 
so that the average of the  magnitude of a tangential vector $\bf u(s)\rm$
 is fixed even if 
the magnitude of tangential vectors  fluctuate. 
This theory provides continuous  crossover formula for the  
extension from the small force limit  to the   large force limit.  In small force limit  the extension $z$ increases  linear    whereas   in large force  limit the  $ f^{-1/2}$ rule is found.  The theory also reproduces quantitatively experimental 
the force-extension curves at high salt concentration.

In this paper, we include   electrostatic interactions in  the  
mean field model of semiflexible chain.
This approach  provides an adequate   
theory of DNA elasticity in  the small force limit.
We compare our results with experiments at  various  salt concentrations.
  
\section{Semiflexible chain under tension}

We model DNA as a semiflexible worm like  chain (WLC) with contour length $L$.
 The chain  can be parameterized  by unit tangent 
vector $\bf u\rm(s) = \partial \bf r\rm(s) / \partial s $, where $\bf r\rm(s)$ 
represents the position vector in three dimensions at the curvilinear position $s$.  The energy cost for bending is characterized by the  persistence length $l_p$.  The tangent  vector for WLC chain satisfies  
 the local   geometric  constraint $\bf u\rm^2(s) = 1$ for all $s$. 
The enforcement of this constraint makes the theory for interacting   WLC
 extremely difficult.  Recently, it has been shown\cite{ha} that  one  can get reliable 
results for a number  of problems  involving   semiflexible chains
 by  replacing the local constraint $\bf u\rm^2(s) = 1$ by a global constraint 
$<\bf u\rm^2(s)> = 1$.  The resulting theory, which in  the 
absence of  interactions  reduces  to the model for semiflexible chains
 proposed  by Lagowski et al.\cite{LNN}, can be  systematically  derived from 
a  functional integral approach.  Here we adopt this mean field 
model to  investigate the    effects of tension.  

The probability distribution  for isolated  WLC  chain, which preserves 
 the global constraint  $<\bf u\rm^2(s)>=1$   is  given by
\begin{equation}
P_o = exp [-\lambda \int_0^L ds \bf u\rm^2(S) -\eta\int_0^L ds (\partial \bf u \rm/\partial s)^2 -b(\bf u\rm^2_L  +\bf u\rm^2_0) ] .
\end{equation} 
The Lagrange multiplier  $\lambda= \frac{3}{2l}$ 
(used to enforce the global constraint 
  $<\bf u\rm^2(s)>=1$) is (roughly)   
inversely proportional to the Kuhn length $l$ of the  semiflexible chain,
 and the  constant  $\eta$ is related to the persistence length $l_p$.
   When the free energy of the non-interacting chain is optimized 
with respect to $\lambda$ and   we obtain  
$\lambda=9/(8l_p)$ and $b=3/4$ .
 These values satisfy the   constraint    $<\bf u\rm^2(s)>=1$.
 
The distribution function of the isolated semiflexible chain under tension  
is given by 
 \begin{equation}
 P_o(f)= exp [- \int_0^L ds \lambda(s)\bf u\rm^2(S) -\eta \int_0^L ds (\partial \bf u \rm/\partial s)^2 + \int \bf f\rm(s) \cdot \bf u\rm(s)-b(\bf u\rm^2_L  +\bf u\rm^2_0) ] 
\end{equation}  
If the applied force is constant,  $\lambda(s)$  has a  uniform  value  for all $s$\cite{ha}.  It  has been  shown that this assumption gives 
a self-consistent solution  to the   stationary condition\cite{ha}.
The optimization  of the free energy, in the presence of a constant force
leads to the modified relation   between $\lambda = 2l/3$ and $\eta$
\begin{equation}
1-\frac{3}{4}\sqrt{\frac{\eta}{ \lambda}} = \frac{f^2}{4 \lambda ^2}.
\label{lambdavseta}
\end{equation} 
The mean square end-to-end distance  of semiflexible chain  under  constant
 tension  can be obtained as 
\begin{eqnarray}
<\bf R\rm^2>&=& \int_0^L \int_0^L < \bf u\rm(s') \cdot \bf u\rm(s'') > ds' ds''\nonumber \\
            &=& l L - l\sqrt{\frac{2 l \eta}{3}}
 ( 1-exp(-\sqrt{\frac{3}{2 l \eta}} L))
+ \frac{l^2 f^2}{9} L^2 
\label{r2}
\end{eqnarray}

\section{ Effects of electrostatic interactions:Interplay of length scales }  

The WLC under  the  influence  of  constant tension yields 
the correct dependence of the 
extension (along the direction of the force)  of $f$ at relatively 
large value of the inverse Debye screening length.  However, precise 
 experiments
on DNA have demonstrated  that the simple elasticity model is inadequate 
to take electrostatic effects into account especially at small  
value of the $f$ and at low salt concentrations. Since DNA is highly charged
 the response of the chain to  tension clearly   depends on the ionic strength
 and to the valence  of the counterions. 
In the presence of the external force there are four important length scales.
They are  the Bjerrum length $l_B$ (= $ e^2/4\pi\epsilon  k_BT$),
the Debye screening length $\kappa^{-1} \equiv r_D$, the persistence length
 $l_p$, and the Pincus length $\xi_P = k_BT/f$\cite{pincus}.
For monovalent ions $\kappa ^2  = 8\pi l_b \rho $ 
where $ \rho $ is the salt(say NaCl) 
concentration.   The persistence length is  $l_p= l_o + l_e$ where $l_o$ 
is the bare persistence length and $l_e$ is the electrostatic contribution.
Although it is well  accepted  that for intrinsically stiff chain 
$l_e \sim \kappa^{-2}$  the dependence of $l_e$ on $\kappa$ for flexible 
chains  depends on a number of factors\cite{barrat&joanny,OSF1,OSF2,semielec,kremer,odijk2,LS}.
 
Since the conformation of DNA subject to tension and salt can change 
from being rod-like to coil-like a proper description requires a detailed 
understanding of the scale dependence of $l_e$.  
The presence of several length scales and the lack of simple theories for 
 variation  of $l_e$ with $\kappa$ make the development of scaling theories 
for elasticity of DNA under tension difficult. 

In certain regimes, however, one can devise a scaling type analysis.
At low salt concentrations the Coulomb repulsion between various charge
residues   leads to stiffening  of the chain. 
Consider the limit  $l_o < r_D \ll L$.  In this case the chain can be modeled as  being  flexible  on scales  larger than $r_D$. Inside the  electrostatic blob
of size $r_D$,
monomers are stretched due to the Coulomb repulsion.
 Now consider the case of small force ($\xi_P > r_D$). 
The mechanical force orients the chain along the tension axis on scales 
larger than $\xi_P$.  The extension is thus given by 
\begin{equation}
z \sim \xi_P (N/N_P)
\label{scaling1}
\end{equation}
where $N_P$ is the number of monomers in a Pincus blob,
 and $N \equiv L/a$ with 
$a$ being the monomer size.
The structure within a Pincus blob is imagined to consists of several 
electrostatic blobs of size $r_D$. 
Within each Pincus blob the mechanical forces are not relevant and
 the arrangement of electrostatic blobs  is dictated   by thermal fluctuations.
Thus,
\begin{equation}
\xi_P \sim r_D (N_P a /r_D)^{\nu}
\label{scaling2}
\end{equation}  
Combining these Eqs. (\ref{scaling1}) and (\ref{scaling2}), we get 
\begin{equation}
z/L \sim \xi_P^{1-1/\nu} r_D^{1/\nu-1}
\end{equation}
where $\nu$ is the usual Flory exponent.
For a highly charged object, the excluded volume interactions are
negligible, and hence it is appropriate to take $\nu = 1/2 $ whence 
we get at small force,
\begin{equation}
z/L \sim (f/k_BT) r_D
\label{blob}
\end{equation}

 The blob picture is not appropriate for intrinsically stiff chains
under a large external force.
This region corresponds to $\xi_P \approx r_D \ll l_p\sim L$.
In this case, the orientation of the tangent vector is correlated 
over scales larger than that determined by Coulomb repulsion or mechanical energy. Thus segments of chain of length $r_D$ are not independent of one another.  In this case, as suggested by Marko and Siggia\cite{marko}, we will show that 
the applied force effectively increases the persistence length of the chain.
For values of $f > f_c = k_BT/(l_o + l_{OSF})$ the chain adopts a highly
stretched conformation.  In the limit of large force 
in high salt concentrations Coulomb interactions are unimportant. 
As discussed previously, the conformation here is entirely  
determined by the competition between 
entropic elasticity and mechanical energy.
With this as the background we develop a detailed theory for dealing with 
electrostatic interactions, at the simple Debye H\"uckel level,
for DNA under tension.

\section{Self-consistent Theory: small force regime }
 In order to take electrostatic effects into account we assume, for simplicity,
that     
 DNA molecule is uniformly charged and each charged  segment of the 
 chain interacts  via screened Coulomb  interactions. 
The Hamiltonian of  an isolated  DNA molecule  consists of  non-Coulomb part
  $H_o$ and  the  electrostatic energy contribution $\Delta H$.
\begin{equation}
H_t = H_o + \Delta H  
\end{equation}
The new probability function $P[\bf u\rm]$ including the electrostatic interaction is 
\begin{equation}
P[\bf u\rm] = P_o[\bf u\rm]exp(\frac{-\Delta H}{k_B T}) 
            = P_o[\bf u\rm]exp[\,- \omega \int^L_0 \int^L_0 ds''ds' 
      \frac{exp(-\kappa|\bf r\rm(s'') - \bf r\rm(s')|)}{|\bf r\rm(s'') - \bf r\rm(s')|} ]
\end{equation}
where $\omega$ is equal to $l_B/A^2$, $A$ is the distance between the charges.

When the  applied force is  smaller than $ f< k_BT/r_D$,  the orientation of 
the tangential vectors are  not correlated  at  scales   $ r > r_D$,
which implies that  the mean square average of angle fluctuation  $<\theta(s)^2>$ is not small.
The chain becomes flexible  on  large scales  and then adopts  
 ``coil like'' conformation, 
although  the chain is still stiff on  small length scales.
We use ``coil-like'' to imply that relatively large excursions from 
rod-like conformation are possible so that the Gaussian approximation,
which would be valid when  $<\theta^2(s)>$  is small, breaks down. 

For this problem the Gaussian approximation employed  to treat the long range 
interaction to the WLC model is not valid\cite{marko}.  This can be seen by deviations from  
experiments in the extension - force curves at small forces.(See Fig.(6)
in ref.\cite{marko}.)
Here we use a self-consistent variational theory to describe the effect of 
small force on the conformations of DNA.
 
In order to estimate the size of the charged DNA under tension we follow the 
 uniform  expansion method introduced by  Edwards and Singh\cite{edward&singh}.
Accordingly,  we write
\begin{equation}
H_t = H_1 + B
\end{equation}
where
\begin{equation}
  B = H_o -H_1 + \Delta H
\end{equation}
Hamiltonian and $H_1$ corresponds to the non-interacting theory in which $l$ is replaced 
by a effective Kuhn length $l_1$ and $\Delta H $  is a  perturbation 
in Hamiltonian.
 The appropriate  value of $l_1$ 
should  satisfy (See Eq.(\ref{r2}).)
\begin{equation}                    
<R^2>= l_1 L - l_1\sqrt{\frac{l_1\,\, l_p}{3}} ( 1-e^{-\sqrt{\frac{3}{l_1\,\, l_p}} L}) + \frac{l_1^2  f^2}{9} L^2. 
\label{selfr2}
\end{equation}
 
For arbitrary choice of $l_1$ , 
 $< R^2 >$ can be rewritten up to first order in  $B$ as  
\begin{equation}
<R^2> = <R^2>_1 -<B R^2>_1 + <B>_1<R^2>_1
\end{equation}  
where $<>_1$ indicates average with weight factor $exp(-\frac{H_1}{k_B T})$.
Self-consistency condition requires that $<B R^2>_1 = <B>_1<R^2>_1$
so  that  $<R^2>$ to first order in $B$  coincides  with that 
 computed  using a reference .
We assume that  the  Kuhn length  $l=2/3\lambda$ 
 will be replaced by $l_1$ by  the coarse graining processes
 in such a way that new parameter $l_1$ satisfies Eq.(\ref{r2}) at given force and  persistence length $l_p$.
Therefore, we have the  following self-consistent equation:
\begin{equation}
(\frac{1}{l}- \frac{1}{l_1} ) l_1^2 [L - \sqrt{\frac{9 l\, l_p}{8}}(1-\exp(-\sqrt{\frac{3}{l_1\,\, l_p}} L)) + \frac{l_1}{2} \exp(-\sqrt{\frac{3}{l_1\,\, l_p}} L) + \frac{2 l_1}{9}(\beta f)^2 L^2] = 
<\Delta z^2(l_1)>_{\Delta H}.
\label{sc} 
\end{equation}
The  right hand side of Eq.(\ref{sc}) is evaluated using 
\begin{equation}
  <\Delta z^2(l_1)>_{\Delta H}= <z^2>-<z^2>_0  =  -\frac{\frac{\partial ^2}{\partial k^2} G(\bf k\rm,L,\bf f\rm)}{G(\bf k\rm,L,\bf f\rm)}|_{\bf k\rm = 0} -<z^2>_0
\end{equation}
where  $<>_o $ indicate the average with weight $e^{-H_o/k_BT}$, and 
$G(\bf r\rm)$ is the  Green  function associated   with the total Hamiltonian
\begin{equation}
G(\bf r\rm, L, \bf f\rm)= \int^{\bf r\rm(L)=\bf r\rm}_{\bf r\rm(0)=\bf 0\rm}
D[\bf r\rm(s)] exp(\frac{-H_t}{k_B T})   
\end{equation}
and its Fourier transform  is 
\begin{equation}
G(\bf k\rm, L, \bf f\rm)= \int d^3 \bf r\rm(s)exp(-i\bf r\rm \cdot  \bf k\rm)G(\bf r\rm,L,\bf f\rm).
\end{equation}
If the applied force is constant along the contour  then 
 $G(\bf k\rm, L, \bf f\rm) = G(\bf k\rm - i\beta  \bf f\rm, L)$ \cite{vilgis}.  We can obtain the mean-square average of the end-to-end distance 
from $G(\bf k\rm, L)$,  
\begin{equation}
 <z^2>=  -\frac{\frac{\partial ^2}{\partial k^2} G(\bf k\rm,L,\bf f\rm)}{G(\bf k\rm,L,\bf f\rm)}|_{\bf k\rm = 0} =
 -\frac{\frac{\partial ^2}{\partial k^2} G(\bf k\rm,L)}{G(\bf k\rm,L)}|_{\bf k\rm = -i\beta \bf f \rm}.
\end{equation}
The correlation function in $k$ space is obtained by performing functional
integral including phase factor $e^{-i\bf k\rm \cdot  \bf r\rm}$.
If we consider the electrostatic interaction as a perturbation, 
$H_t = H_o + \Delta H$, $\Delta H = V_{DH}(r)  = l_B e^{-\kappa r} /r $.  
\begin{equation}
G(\bf r\rm, L) 
= \int \frac{d^3\bf k\rm}{(2 \pi)^3} <e^{ikr}>_o -\sum_{n} \frac{\beta}{n!} 
<(\Delta H  e^{ikr})^n>_o  
\end{equation}
We can write  the correlation function  $G(k) \approx G_o(k,L)
\frac{1}{n!}\sum_{n=0} G_1(k,L)\approx G_o(k,L)\exp(G_1(k,L))$
 under the Gaussian  approximation.
\begin{eqnarray}
G_o(k,L)G_1(k,L) &=& < V_{DH}  \,e^{ikr} >_o   \nonumber\\ 
             &=& \frac{1}{(2 \pi)^3}\frac{\omega}{2}
\int ds''\int ds' \int d^3 q
    \frac{1}{q^2 + \kappa^2} < e^{i\bf q \rm  \cdot (\bf r\rm(s'') -\bf  r\rm(s'))  
+ i\bf k \rm  \cdot (\bf r\rm(L) -\bf  r\rm(0))}  > \nonumber \\
&=& exp[-\frac{k^2 l}{6 a}(a L -1+ e^{-aL})] \times\nonumber \\
  &&\mbox{}\frac{\omega \sqrt{\pi^{3}}}{16 \kappa^2 } \int ds \int ds' \int_0^{\infty} d\alpha e^{-\alpha}(\frac{\pi}{g_1 l_1 + \alpha/\kappa^2})^{\frac{3}{2}}
 exp(\frac{l_1^2 g_2^2 k^2}{g_1 l_1 + \alpha/\kappa^2})
\label{sc2}  
\end{eqnarray}
with
\begin{eqnarray}
a &=& (\frac{3}{2 l \eta})^{\frac{1}{2}} \nonumber\\
g_1(s''-s')&=& \frac{1}{6a} ( a(s''-s') -1 +  e^{-a (s''-s')}) \nonumber \\ 
g_2(s''-s')&=& \frac{1}{6a} ( a(s''-s') -1 -2 e^{a L /2} \sinh(s''-s')a/2 \cosh(L-s''-s')a/2)\nonumber 
\end{eqnarray}
where  the wave vector q is associated with the  momentum transfer
 via electrostatic 
interaction and $\frac{1}{q^2 + k^2} $ is the Fourier transform of 
$V_{DH} (\bf r\rm(s'') - \bf r\rm(s') )$. We have  introduced 
the dummy parameter
$\alpha$ in order to replace three dimensional  integral 
with respect to   $\bf q\rm$  to a one dimensional  integral  in $ w$,
 The mean square average of  the end-to-end distance is 
\begin{equation}
 <\Delta z^2>_{\Delta H}= <z^2>_0  +
 (2 \frac{ G_o'(\bf k\rm,L) G_1'(\bf k\rm,L)}{G_o(\bf k\rm, L)} + 
G_1''(\bf k\rm, L) )| _{ \bf k\rm = -i\beta\bf f\rm}
\label{delz2}
\end{equation}
We solve Eq.(\ref{sc}) together with  Eqs.(\ref{sc2}) and (\ref{delz2}) 
iteratively to find  a new 
coarse grained Kuhn length $l_1$,  which is related to the persistence
 length via Eq.(\ref{lambdavseta}).  
The correction to $<z^2>$ due to the electrostatic interaction
 is  expected  to be always  positive  since   
electrostatic interactions stiffen the chain.
 Therefore  we  would expect  
the effective Kuhn length  $l_1 $, which varies with salt concentration, to be larger than $l$.   
 
In order to analyze the experimental measurements at small force using our theory we need  the parameters L (the contour length), $l_o$ (the intrinsic persistence length of DNA), and the effective linear  charge density $1/A$.  The 
values  of L and $l_o$
may be  obtained by fitting the force-extension curve to the data of 
Smith et al.\cite{smith2} at the monovalent salt concentration of $10 mM$ NaCl
($\kappa^{-1} \approx$ 3.2 nm) using WLC model\cite{doi&edwards}. 
For this condition  the WLC gives an excellent description of the data because 
the electrostatic interactions are negligible\cite{marko,ha}. 
The best fit is obtained with $L = 32.7 \mu m$ and with the intrinsic 
persistence length $l_o = 53 nm$.
 The effective charge density $1/A$ still remains  free parameter.  
In this 
paper,  we choose that $1/A$  in order to find the best fit with experiments.
We will discuss the effect of ion condensation in the following section.  
We use the Bjerrum length $l_B = 0.7 nm $ in water at room temperature with dielectric
constant $\epsilon = 80$.  

Before we present the results of the force-extension it is useful to characterize the variation of the electrostatic persistence length of DNA with $\kappa$.
Here $l_e = l_p -l_o$, $l_p = 3l_1/4$ and $l_o = 53 nm$.
In Fig.(\ref{endtoend}) we plot $l_e$ as a function of $\kappa$.  
It is clear that there are two distinct scaling  regimes of behavior.
\begin{eqnarray} 
             l_e \sim &\kappa^{-1} &  \kappa < \kappa_{x} \\ \nonumber
                      &\kappa^{-2} &  \kappa  > \kappa_{x}
\end{eqnarray}
where $\kappa_{x}\approx 0.4 nm^{-1} $ is roughly the crossover value.
This behavior is in accord with recent  theoretical predictions\cite{semielec,netz}.
  The crossover,
at least for this DNA, occurs when  $l_e \approx l_o$\cite{kremer}, which is consistent with
the condition  $l_{OSF} \approx l_o$.
In  the inset to  Fig.(\ref{endtoend}), we  plot 
the salt concentration  dependence of the radius of gyration of DNA.
   In the  low  salt  concentration  region, 
$\kappa < 0.1 nm^{-1} $, the radius of gyration $R_g$ varies like  
 $\kappa^{-1/2}$  and 
 for high salt concentration, $R_g$ shows  little $\kappa$ 
dependence. The  $\kappa^{-1/2}$ variation, shown in the inset to Fig.(\ref{endtoend})  implies $\nu = 1/2$  for  $R_g$. 
The implication of this calculation, for our purpose, is that in the small 
$\kappa$ regime $l_e > l_o$ and as a result the effects of electrostatic 
interaction dominate at scales of $l_e$.  Thus, a more detailed theory 
described here is required to 
describe the  small force behavior of DNA.  

In Fig.(\ref{force_extension}), we plot force-extension  curves for 
DNA molecule  at  $ 10 mM $ $  Na^{+}$, $ 1 mM $ $  Na^{+}$,
$ 0.1 mM $ $  Na^{+}$  ion concentration respectively.
The large force fits are done by the  calculation of  the angle fluctuations of
 the tangential vector  using $L = 32.7 \mu m$
and $l_o = 53 nm$.    (See next section.)
 The small force fits are done 
with self-consistent mean field approximation
using the same values  for the parameters $L$ and $ l_o$.

The choice of $A$, for which direct measurements are not available, requires 
explanation.   
We choose the  effective charge density $1/A$ as  $ 2.5 nm^{-1}$, $ 1.4 nm^{-1}$
and $1.0 nm^{-1}$ for $ 10 mM $ $  Na^{+}$, $ 1 mM $ $  Na^{+}$,
$ 0.1 mM $ $  Na^{+}$ ion concentration respectively. These values  give
 the best fit to the data.
At large salt concentration $10mM Na^{+}$, the  interaction range (Debye radius)   $r_D = 0.3 nm$  is smaller than the size of each base of DNA.  We conclude that counterion  condensation is not relevant in this concentration range.
If every base pair  carries a charge $-1e$, 
 the  linear charge density will be 
$2.94 nm^{-1}$ since the size of base pair is approximately  $0.34 nm$.    
The choice of our  effective charge density $1/A = 2.5nm^{-1}$ at $10mM$ 
NaCl concentration  indicates that most of  the counterions  dissociate from 
the monomers on the backbone.
At smaller salt concentration, namely $1 mM Na^{+}$ and $0.1 mM Na^{+}$,
the interaction range exceeds the Bjerrum length  of aqueous solution,  
 $l_B = 0.7 nm$.
We expect counterions  are condensed in the vicinity of DNA, which leads to 
a reduction in the effective charge density.  According to the Manning condensation
theory\cite{manning,BJ},  the charge density larger than one per Bjerrum length  leads to 
counterion condensation.  Therefore we choose $1/A = 1.4 nm^{-1} \approx 1/l_B$.
We used smaller value of $1/A$ for $0.1mM$ concentration.  However, 
all the data points from experiment correspond to the rod-like conformation 
even in the  small force regime.
We expect   the self-consistent theory to be  valid  even at very  smaller
values of  force, 
$f\leq 10^{-2} pN$, and the approximate choice  of $1/A$ is then  of  the order of $1/l_B$.

It is clear from Fig.(\ref{force_extension}) 
 that there are two different regimes  in the chain elasticity as the  
applied force increases.  They  correspond to 
 the   ``coil-like''    and rod-like  conformations respectively.  
At a given ionic concentration, we can observe a  plateau 
in the intermediate  force regime.  The validity of self-consistent mean field 
calculation is limited before the onset of  plateau, where  
``coil-like'' conformations dominate.  Our  calculations  provide 
the correct estimation 
of the chain extension   when  the extension $z$ is much smaller
 than  the total contour length $L$ i.e. $z/L < 0.5$. 
The  mean square average $<R^2>^{-1/2}$ with zero force is proportional 
to $\kappa^{-1/2}$ 
and increases as $f\kappa^{-1} $ as the magnitude
of force increases.  These results   agree  with experiments 
and the blob picture (see Eq.(\ref{blob})) in the small force limit.
At  smaller ionic concentrations, the position of the  plateau moves to  the  
smaller  values of force.
  In the following section, we 
discuss the force-extension curves for  a semiflexible chain in a rod-like 
conformation.

\section{Electrostatic effects on stretching stiff DNA:
Large force regime}
In this section, we  apply the   functional integral method to obtain 
the force-extension relation  of an intrinsically rigid  chain
  in  the  limit of large force.  
This regime was considered  by Marko and Siggia\cite{marko} who noted  that the effects 
of electrostatic interactions can be absorbed into an effective scale 
dependent persistence length.  Here, we provide a derivation of this result
using a functional integral approach. 
  
It appears that when a  large force is applied to the chain, the segments  of 
the chain  do not interfere with each other geometrically,
because of tension  induced  stiffening occurs  on scales from $\xi_P = k_BT/f$.
 If this  assumption is valid   the   influence of the  electrostatic interaction 
 can  be treated simply   by  replacing 
the  persistence length $l_p$  by an effective persistence  length $l_p^{eff}= l_o + l_{OSF}$.
With this replacement the force-extension  curve can be easily calculated.
Since the backbone is intrinsically stiff it follows that 
the fluctuation in  angle $\theta(s)$, $\cos\theta(s)= \bf u\rm(s)\cdot\bf u\rm(0)$,  is small, we can expand $cos(\theta(s)) \approx 1+\frac{1}{2!} \theta^2(s) + \frac{1}{3!}\theta^3(s) + ... $  If we consider only fluctuations of the angle,
  the expansion with respect to 
 $\theta(s)$  guarantees  the constraint of  $|\bf u\rm(s)^2| =1$.     
The Hamiltonian of the system can be written in terms of $\theta(s)$ as,
\begin{eqnarray}
\frac{H}{k_B T}& \propto & \frac{l_p}{2}\int^L_0(\frac{\partial \bf u\rm(s)}{\partial s})^2 ds + \frac{\omega}{2}\int^L_0  ds'' 
\int^L_0 ds' \theta(s'')\theta(s') G^{-1}(s''-s') - \int^L_0 \frac{|\bf f\rm(s)|}{k_B T} cos\theta(s) ds  \nonumber \\
&=& \frac{1}{2} \int^{\infty} _{-\infty} \tilde{\theta}^2(q)Q^{-1}(q) dq
\label{eql}
\end{eqnarray}
\begin{equation}
\tilde{Q}^{-1}(q) = \tilde{G}^{-1}(q) + l_p q^2+ \frac{\bf f \rm}{k_B T}.   
\end{equation}
where  $\tilde{\theta}(q) = \sqrt{\frac{1}{2 \pi}} \int^{\infty}_{-\infty} 
e^{iqs}\theta(s) ds$, and in the limit of $q/\kappa \ll 1$,
\begin{equation}
\tilde{G}^{-1}(q)= \frac{\omega}{2}[ (1+\frac{1}{q^2 \kappa^{-2}})
 ln(1+ q^2 \kappa^{-2}) -1]
         \approx  \frac{\omega (q/\kappa)^2}{4} - \frac{\omega}{6}(q/\kappa)^4
+  O((q/\kappa)^6).
\end{equation} 
 Therefore, the effective persistence length  $l_p^{eff}$ can be identified 
with  the coefficient of the  quadratic term:  $l_p^{eff} = l_p + l_{OSF}$, $l_{OSF} =  l_B /4 \kappa^2 A^2 $.
  
The generating function $Z$  for the theory given in Eq.(\ref{eql}) is  
\begin{equation}
Z \propto \int D[\theta] exp[-\frac{\omega}{2}\int^L_0 \int^L_0 \theta(s'') G^{-1}(s''-s')\theta(s') ds'' ds' + \int^L_0 \frac{ |\bf f\rm|}{k_B T} \theta(s) ds].  
\end{equation}
The ratio of the extension to the total length  $<z/L>$ can be obtained 
from the generating function $Z$
\begin{equation}
<\frac{z}{L}>  = 1-\frac{<\theta^2(s)>}{2} =  1- \frac{1}{2} \frac{\partial}{\partial f} \frac{\partial}{\partial f} Z  
\label{extension}
\end{equation}
where the mean square value of $\theta(s)$ can be found from $\tilde{Q}^{-1}(k)$, 
\begin{equation}
<\theta^2(s)>  = \frac{1}{2 \pi}\int^{\infty}_{-\infty}dk \int^{\infty}_{-\infty}dk'                   < \theta(k) \theta(k')> e^{is(k-k')}
               = \frac{1}{2 \pi} \int^{\infty}_{-\infty}dk e^{2isk}Q(k).
\end{equation}
In Fig.(\ref{force_extension})  we show 
$z/L$ (Eq.\ref{extension}) as a function  of force.   Our results (See Fig.(\ref{force_extension}))  are   
 in  very good agreement with experimental results 
of Smith et al.\cite{smith2} for all salt concentrations.  
 In the large force regime,  we find  $<z/L>  \propto  -1/\sqrt{f}$
which, of course implies, that is regime   DNA does behave as WLC\cite{marko,ha}.    
It is not surprising that the theoretical results for large forces 
 start deviating 
when $z/L < 0.4$, and dramatically depart from the experimental results 
and when $z/L < 0.2$ 
``coil-like'' conformations dominate  at small forces.  This    suggests that 
  for small
 force regime a more elaborate theory, such as the one presented in the previous section, is required.

\section{Conclusions}
In this paper we have considered the effects of electrostatic interactions 
on the stretching of DNA. It is already well established that the simple 
elastic model at high salt concentration gives an excellent description of the 
response of DNA to tension.  Furthermore,  it is clear that at small 
values of the concentration of salt and at small force ``coil-like'' conformation become important and an elaborate theory  is required.  Here we have shown 
that when the  applied  force is small,  the long range interaction in DNA 
molecule can be properly treated by  self-consistent variational  mean field approximation.
Our theory  gives excellent agreement with experimental results in this regime.
The  larger  the ionic strength, the larger the overlap range  in the cross over regime  from the mean field 
calculation (coil conformation)  to the WLC  limit (stretched conformation).  
If  the screening length $r_D$ is large compared to the total contour length $L$ ($r_D/L \sim 1$),
  then  electrostatic interactions dominate at all scales so that the chain 
is effectively stiffened  even when the applied force is small.
In this case, the  self-consistent theory is valid  only 
in the limit of very small force. 

One of the limitations of treating the electrostatic interactions, even at the 
primitive Debye-H\"uckel level, is that there is no easy way to choose the 
linear charge density $1/A$.  The value of $A$ is essentially controlled 
by counterion condensation effects.  Since the conformation of DNA 
changes upon addition of salt dramatically a proper theory 
of describing DNA elasticity should include fluctuations due to counterion 
condensation.   In the absence of such a theory we have used physical arguments to choose a value for $A$.  A more elaborate theory that would treat 
fluctuations on small scales of the value of $A$ as well as larger length scale is required to obtain the  line density independently.

It appears that in the limit of small $f$ the self-consistent  variational theory may be adequate for all values for $\kappa$.  For large value of $\kappa$ and $f$ the effects of electrostatic interaction may be treated using such a scale
dependent persistence length\cite{marko}.
It would be desirable to have a unified theory that can treat both  
regimes, and hence the case of intermediate values of $f$ and $\kappa$.
Such a theory would require using a more elaborate variational Hamiltonian,  perhaps,  
similar to the ones used recently to treat the persistence length of 
polyelectrolyte chains\cite{netz}.

Despite  the  success of the theory outlined here it is worth pointing out 
that the WLC can only explain the force-extension curves when the counterion
is monovalent.  It is only  in the presence of monovalent ions that the  
electrostatic persistence length of DNA displays  the  well accepted dependence on $\kappa$.  Baumann et al.\cite{baumann}
have found that multivalent salt ions (counterions) have dramatically different effects on persistence length of DNA.  By performing force measurement 
using laser tweezers they discovered  that not only  valence  but also 
the shape of counterions  profoundly affect  elasticity of single DNA molecules.
This observations and other findings  by Baumann et al.\cite{baumann} clearly suggest  that theories 
and simulations that go beyond the simple  Debye-H\"uckel theory will be required  to provide a complete description of the response  of DNA to external tension.
\\

\bf Acknowledgments \rm
This work was supported in part by a grant from the National Science Foundation through  the grant number NSF CHE 96-29845.

\begin{figure}
\caption{ The electrostatic contribution
 to the persistence length $l_{e}$ as a function of $\kappa$.
It is clear that $l_e$ exhibits    
  $\kappa^{-1}$ and $\kappa^{-2}$ dependence 
in low and high salt concentration  respectively.  
The  inset shows  the  dependence of the  radius of gyration  on the salt concentration, which is  calculated by solving self-consistent Eq.(\ref{sc}).
The value of parameters are  $L = 32.7 nm$, $l_o= 53 nm$, $1/A = 1.40 nm^{-1}$.}
\label{endtoend}
\end{figure}
\begin{figure}
\caption{Comparison  between theoretical and experimental force-extension 
curves  for  DNA at various salt concentrations.
 (a) $10 mM Na^{+}$, (b)$ 1 mM Na^{+}$, (c)$ 0.1 mM Na^{+}$
The symbols denote   experimental results,  
and the    dashed lines  represent 
the functional integral estimation for high force limit.   
The  force vs. extension curves  for small force regime (solid lines)
 are  calculated  by numerically solving the  self-consistent equation. 
In panel (c) the solid line is calculated with $A = 1.0 nm$ while 
the dot-dashed  line is computed with $A=1.1nm$. 
In all cases $L= 32.7 \mu m$ and $l_o = 53 nm$.
}
\label{force_extension}
\end{figure}

\end{document}